# Silicon Photonics DWDM NLFT Soliton Transmitter

J. Koch[(1*)], A. Moscoso-Mártir[(2)], J. Müller[(2)], F. Merget[(2)], S. Pachnicke[(1)], J. Witzens[(2)]

(1) Chair of Communications, Technical Faculty, Kiel University, Kaiserstraße 2, 24103 Kiel, Germany, jonas.koch@tf.uni-kiel.de
(2) Institute of Integrated Photonics, Faculty of Electrical Engineering and Information Technology, RWTH Aachen University, Campus Blvd. 73, 52074 Aachen, Germany

*Abstract*— We investigate the transmission of densely multiplexed solitons using a photonic integrated chip and the nonlinear Fourier-transform and analyze required launch conditions, the effect of (de-)multiplexing and noise on the nonlinear spectrum, and equalization techniques that can be used to enhance the transmission performance.

## I. Introduction

Demands on modern communication systems for higher data rates are pushing the capacity limits of currently deployed fiber. One way to increase the maximum data rate is to increase the dimension of modulation formats. However, this also requires a higher signal-to-noise ratio (SNR), which leads to the use of higher signal powers that, in turn, introduce nonlinear distortions inside the fiber. To address these challenges, the nonlinear Fourier transform (NFT/NLFT) has been proposed as a means to expand the limits of usable power per fiber without incurring the aforementioned nonlinear distortions. The NFT is a mathematical tool that linearizes the transmission through the fiber optical channel and enables fiber-optical communications with high signal powers [1]-[3]. A main drawback of this technique stems from the complexity of creating higher-order solitons by an inverse NFT (INFT) at the transmitter, which can be demanding in regards to hardware requirements, such as spectral bandwidth, resolution and sampling rate of the digital-to-analog converters (DACs) [4]. Furthermore, using the NFT technique, in principle all WDM channels have to be processed jointly to accurately describe the propagation of the signal through the fiber, so that multi-channel effects such as four-wave-mixing or cross-phase modulation can be fully compensated.

This paper tackles both of the aforementioned problems by shifting the higher-order soliton creation (soliton merging) into the optical domain [5]. This way, first-order (fundamental) solitons need to be generated only, which require a much lower bandwidth and lower sampling rate. Afterwards, these first-order solitons are modulated onto multiple optical carriers and are multiplexed by means of an optical programmable network realized as a photonic integrated circuit (PIC). Dense multiplexing of solitons into higher-order pulses constrains time and frequency spacings between them. Consequently, this technique imposes some limitations regarding the time and frequency spacings that need to be fulfilled so that soliton merging can be achieved. These limitations are carefully modeled in this paper and their impact on WDM systems based on merged solitons is analyzed.

This paper is organized as follows. A short introduction into the NFT is given in Section II. The third section studies the boundaries of linear soliton superposition for two and four solitons. Section IV discusses different approaches to equalize the received signals. In Section V, the PIC used for tight multiplexing of the solitons is presented and its power budget is calculated. Simulation setups and corresponding simulation results are discussed in Section VI. Finally, conclusions are drawn in Section VII.

## II. The nonlinear Fourier-transform

In the well-known (linear) Fourier transform, complex exponential functions $e^{j\omega t}$ are considered eigenfunctions. After the transform, a signal $q(t)$ is thus decomposed into a superposition of complex exponential functions, which in turn stand for harmonic oscillations. In other words, the linear Fourier transform splits a signal into harmonic wave solutions and represents them according to their frequency. In the corresponding frequency spectrum, the power values of the individual frequency components are plotted.

In contrast to the linear Fourier transform, that describes the signal as a single continuous spectrum, the NFT describes the signal as a combination of a continuous and a discrete nonlinear spectrum, that respectively correspond to dispersive and non-dispersive (solitonic) waves. In the NFT, angular frequencies, as defined for the linear Fourier transform, are replaced by the eigenvalues of an associated linear scattering problem [1],[2]. Both NFT spectra evolve through the nonlinear channel given by an amplified (and thus loss-less) optical fiber as if it were a linear medium, in that the associated coefficients simply accrue an additional phase as for a harmonic oscillation propagating though a linear medium, and, in the case of the discrete spectrum that corresponds to complex-valued eigenvalues, a change in amplitude that represents the timing of the soliton's envelope function relative to a delayed-time reference. This enables transmission in a nonlinear power regime without requiring the computation of e.g. nonlinear backpropagation for equalization purposes [1],[2].

The continuous NFT eigenvalues $\xi \in \mathbb{R}$ can be found on the real axis inside the bandwidth of the signal and are analogous to frequencies in the linear case, as they coincide in the limit of low signal power, with $2\xi$ equal to the linear angular frequency. However, the discrete eigenvalues $\lambda_m$, with $m = 1 \ldots N$, are a finite number of complex-valued numbers in the upper half complex plane with a positive imaginary part. Each describe a soliton in the time domain. By using an INFT, a set number of discrete eigenvalues can be modulated to build a higher-order soliton, which can reach very high signal powers [1]-[3].

In order to modulate the NFT spectrum, the NFT coefficients $\hat{q}(\xi) = b(\xi)/a(\xi)$ and $\tilde{q}(\lambda_m) = b(\lambda_m)/a'(\lambda_m)$, with $\tilde{q}$ the residue of $\hat{q}$ at $\lambda_m$, associated to each of the eigenvalues are modulated (with functions $a$ and $b$ defined in [1],[2]). The coefficients $\hat{q}(\xi)$ are analogous to the field amplitudes associated to a linear spectrum and reduce to the same in the limit of low signal power. Similarly, the phase of $\tilde{q}(\lambda_m)$ and $2\text{Re}\{\lambda_m\}$ describe the phase and the center frequency of a soliton, with a direct correspondence in the case of an isolated, fundamental soliton. $2\text{Im}\{\lambda_m\}$ and $|\tilde{q}(\lambda_m)|$ describe the amplitude $A_m$ and timing $T_m$ of the soliton's envelope function [1],[2], with $A_m = 2\text{Im}\{\lambda_m\}$ and $T_m = \log(||\tilde{q}(\lambda_m)||)/2\text{Im}\{\lambda_m\}$ in the case of an isolated, fundamental soliton. The correspondence between NFT parameters and the spectrum obtained from the linear Fourier transform becomes more complex in the case of higher order solitons or colliding fundamental solitons. For example, during the collision of two fundamental solitons of different center frequencies, their center frequencies transiently shift as a consequence of cross-phase modulation (XPM) [6], even though their respective NFT eigenvalues remain unchanged. This is an aspect that will play a role in the receiver impairments discussed in the following [7].

In general, the resulting signal after INFT can be a combination of discrete and continuous spectra, including purely solitonic or purely continuous (dispersive) spectra, which satisfy a generalization of Parseval's identity [1]

$$E = \int_{-\infty}^{\infty} |q(t)|^2 \, dt = E_{\text{disc}} + E_{\text{cont}} \quad (1.1)$$

$$E_{\text{disc}} = 4 \sum_{m=1}^{N} \text{Im}\{\lambda_m\} \quad (1.2)$$

$$E_{\text{cont}} = \frac{1}{\pi} \int_{-\infty}^{\infty} \log(1 + |\hat{q}(\xi)|^2) d\xi \quad (1.3)$$

wherein $E$ is the total signal energy, which is decomposed into the energies of the discrete ($E_{\text{disc}}$) and continuous ($E_{\text{cont}}$) spectra. The energy of the discrete spectrum is linearly dependent on the solitons' amplitudes and – as a consequence – the time-domain pulse widths are also depending on them [8].

## III. BOUNDARIES OF LINEAR SOLITON SUPERPOSITION

Several challenges are associated with the multiplexing of fundamental solitons at the transmitter. One is a purely practical one and arises from partially overlapping soliton spectra (for higher spectral efficiency), that precludes the use of conventional wavelength division multiplexers, as they would lead to a truncation of the spectra. This is addressed by multiplexing with a custom PIC and is further described in Section V. Another challenge arises at a more theoretical level, in that ideal multiplexing can also be seen as a linear superposition of the incoming fields. This operation is not part of regular fiber propagation, there is thus no a-priori guarantee that eigenvalues and associated NFT coefficients are being conserved after multiplexing. In other words, a superposition of fields in the linear domain does not correspond to a superposition of eigenvalues in the NFT domain, as opposed to what happens with the linear Fourier transform.

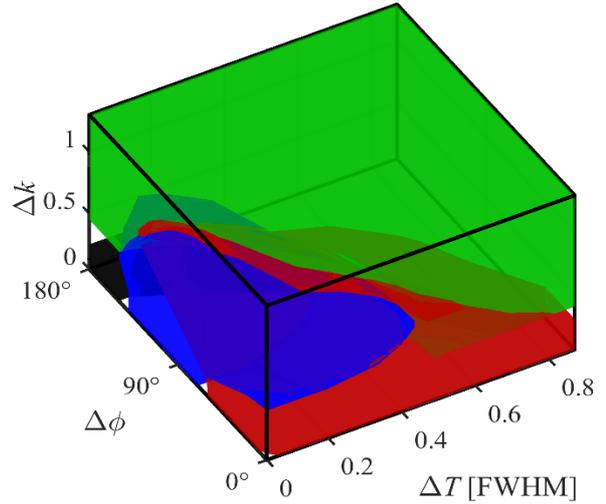

Fig. 1. Three dimensional representation of the outcome of two-eigenvalue combinations depending on $\Delta k$, $\Delta T$ and $\Delta\phi$. Green stands for merging, red for co-propagation and blue for fusion.

Superposing multiple solitons can thus lead to a modification of the solitons' propagation properties, as well as to a partial conversion into dispersive waves, as will be shown in Subsections III.A and III.B. These effects depend on the differences between the eigenvalues' properties, such as their relative positions in the complex plane and the values of the corresponding NFT coefficients $\tilde{q}(\lambda_m)$ [4],[9]. To fully conserve the value of the initial eigenvalues $\lambda_{m,n}$ when merging two fundamental solitons, either one of two boundary conditions relating to timing or center frequency have to be met. When the time difference is chosen to be very high, propagation of each eigenvalue after merging resembles that of a solitary fundamental soliton transmission using only one carrier with

$$\lim_{\Delta T_{m,n} \to \infty} (\Delta \check{\lambda}_{m,n}) = 0 \quad (2.1)$$

wherein $\Delta T_{m,n} = |T_m(\tilde{q}_m, \text{Im}\{\lambda_m\}) - T_n(\tilde{q}_n, \text{Im}\{\lambda_n\})|$ is the time difference between the center positions $T_{m,n}$ of the solitons with indices $m$ and $n$, and $\Delta\check{\lambda}_{m,n}$ stands for the deviation, after superposition, of eigenvalue $\lambda_m$ from its initial value, depending on eigenvalue $\lambda_n$ and their respective NFT coefficients $\tilde{q}_{m,n}$.

The second condition is the frequency spacing in Fourier domain, which is determined by the difference between the eigenvalues' real parts. If this value is high, a WDM-

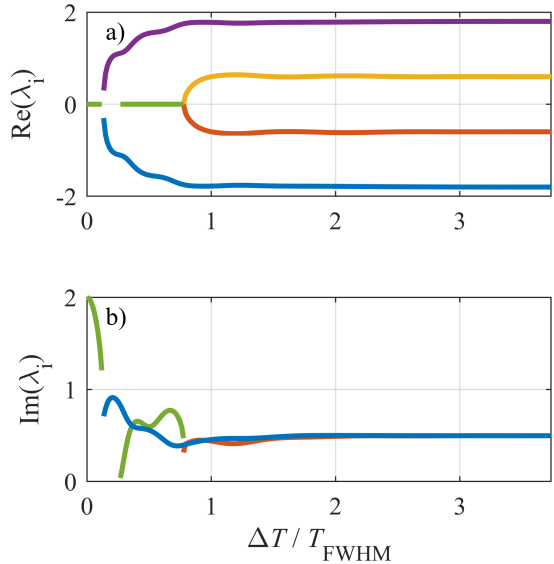

Fig. 2. Exemplary eigenvalue decomposition after superposition of four eigenvalues as a function of $\Delta T$. (a) Real value and (b) imaginary value.

like transmission can be assumed after merging, for which we again obtain

$$\lim_{\Delta f_{m,n} \to \infty} (\Delta \check{\lambda}_{m,n}) = 0 \quad (2.2)$$

wherein $\Delta f_{m,n} = |f_m(Re\{\lambda_m\}) - f_n(Re\{\lambda_n\})|$ is the difference of the center frequencies of the two solitons.

In the above, times, lengths and power levels were expressed as dimensionless multiples of characteristic scales $T_0$, $L_0$ and $P_0$, with $L_0 = T_0^2/|\beta_2|$, $P_0 = 1/\gamma L_0$, $\beta_2$ the characteristic chromatic dispersion of the fiber (in ps$^2$/km) and $\gamma$ the Kerr coefficient of the fiber (in W$^{-1}$km$^{-1}$). In the following, the imaginary parts of the eigenvalues of all solitons are fixed to $Im\{\lambda\} = 0.5$ and the time-domain full width at half maximum (FWHM) of all initially generated fundamental solitons is assumed to be identical and set to $T_{FWHM} = 1.763 \cdot T_0$.

This leads to the following simplifications:

$$\Delta f_{m,n} = \frac{1}{\pi T_0}|Re\{\lambda_m\} - Re\{\lambda_n\}| = \frac{1}{\pi T_0}\Delta k \quad (3.1)$$

$$\Delta T_{m,n} = \log\left(\frac{|\tilde{q}_m|}{|\tilde{q}_n|}\right) T_0, \quad (3.2)$$

where $\Delta f_{m,n}$ and $\Delta T_{m,n}$ are expressed in physical units and $\Delta k$ is the difference between the real parts of the eigenvalues.

*A. Two eigenvalues*

If two solitons are multiplexed with close timing and center-frequencies, both solitons' eigenvalues and their associated NFT coefficients experience changes. Furthermore, part of the energy can be transferred into the dispersive (continuous) spectrum, fulfilling energy conservation as given by Eq. (1.1). This is then accompanied by a reduction of the imaginary parts of the eigenvalues. Besides $\Delta k$ and $\Delta T$, these changes also depend on $\Delta \phi$, the phase difference between the two solitons' NFT coefficients at the time of the superposition. A thorough investigation of soliton superposition is reported in [4],[9]. In summary, three types of eigenvalue combinations can occur: i) co-propagation, which corresponds to two eigenvalues with different imaginary parts and the same real part and results in higher-order soliton breathers; ii) fusion, which is indicated by only one remaining discrete eigenvalue with a higher imaginary value than the original ones and corresponds to a single fundamental soliton, with the possibility of some power being lost to dispersive waves; and iii) merging, which corresponds to two eigenvalues with the same imaginary value and different real parts and corresponds to independently propagating fundamental solitons (also referred to as soliton repulsion [9]).

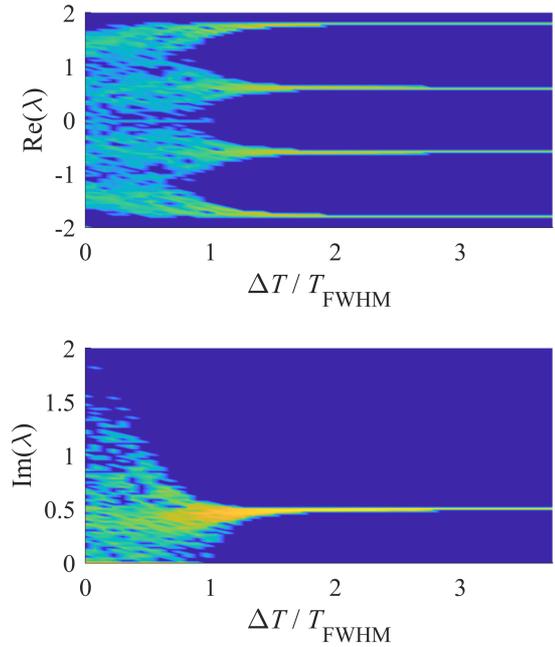

Fig. 3. Eigenvalue positions, after superposition, as a function of $\Delta T$ for all possible combinations resulting from QPSK modulation, plotted as a color-coded histogram.

The last case, i.e., separate eigenvalues with different real parts after superposition, resembles WDM transmission and is the desired outcome in the following sections for straightforward data demodulation. The different post-superposition outcomes are summarized in Fig. 1. From this dataset, it can for example be seen that if soliton merging with arbitrary phase modulation of the individual NFT coefficients is desired, $\Delta k > 0.75$ is required at an exemplarily chosen value of $\Delta T/T_{FWHM} = 0.6$. If $\Delta k < 0.75$, co-propagation results when the phase difference of the superposed solitons is below a certain threshold, thus not allowing arbitrary phase modulation with pure soliton merging. Furthermore, if merging is implemented with soliton parameters close to the co-propagation boundary, the post-merging eigenvalues and NFT coefficients can strongly deviate from their initial values. Hence, the combined lower boundary of $\Delta T$ and $\Delta k$ is further shifted upwards, as described in [4].

*B. Expansion to four eigenvalues*

If more than two eigenvalues are linearly superimposed, channels further interfere with each other, leading to combinations of the effects described in Subsection III.A. However, for a large number of soliton channels, channels far from each other asymptotically reach condition (2.2). For the example shown in Fig. 2, a fixed channel spacing of $\Delta k = 1.2$ was chosen in order to comply with the frequency spacing used in the following sections of this paper. The remaining free parameters are thus $\Delta T$ and $\Delta\phi$. However, since we will apply QPSK modulation to the eigenvalues' NFT coefficients, $\Delta\phi \in [0, \pm 90°, 180°]$ was used. The post-superposition values of the four eigenvalues are shown in Fig. 2 as a function $\Delta T$, applied to each of the pulse pairs, i.e., the pulses are assumed to be equidistant in their transmission window. Here, the real parts of the eigenvalues were set to $\text{Re}\{\lambda\} = [-1.8, -0.6, 0.6, 1.8]$ and $\Delta\phi$ set to 0 between all neighboring channels.

As can be seen in Fig. 2(a), a fusion of all four eigenvalues into one high energy eigenvalue occurs for $0 \leq \Delta T/T_{FWHM} < 0.135$. Besides, a noticeable decrease of the eigenvalues' imaginary value (or, equivalently, a shift of the energy into the continuous spectrum) is shown in Fig. 2(b) as $\Delta T/T_{FWHM}$ approaches the boundary at 0.135. Afterwards, a combination of fusion and merging occurs. First, four initial eigenvalues fuse into two eigenvalues with different real parts for $0.135 < \Delta T/T_{FWHM} < 0.27$. Next, a third eigenvalue arises for $0.27 < \Delta T/T_{FWHM} < 0.78$, which finally splits into two eigenvalues with different real parts for $\Delta T/T_{FWHM} > 0.78$. From here on, pure merging occurs, with the conservation of four eigenvalues with four different real parts. However, one can see that the eigenvalues are not on their desired, initial spectral positions until $\Delta T/T_{FWHM}$ is roughly set to 1, i.e., $\Delta T \approx 1.76$ in the dimensionless system of units defined above. The simulation from Fig. 2 has been repeated for the evaluation of post-superposition eigenvalue positions, as a function of $\Delta T$, for all 256 possible combinations of the four QPSK-modulated NFT coefficients ($\Delta\phi \in [0, \pm 90°, 180°]$). This is shown in Fig. 3 as a color-coded histogram. It makes apparent that a longer $\Delta T$ above 2.7 ($\Delta T/T_{FWHM} > 1.5$) is needed for all eigenvalues to reach their desired position for all $\Delta\phi$ combinations after linear superposition.

## IV. EQUALIZATION TECHNIQUES

The equalization implemented in this paper is set up to reduce the effect of three categories of distortions. The first distortion is caused by noise. Since noise leads to nonlinear interactions in the nonlinear Fourier domain, the effect of noise is non-trivial. Noise can cause correlated deviations of the eigenvalues and their NFT coefficients [10], that can be exploited for noise reduction. If only $b(\lambda)$ is considered instead of the ratio of the $b(\lambda)$ and $a'(\lambda)$ coefficients ($\tilde{q}(\lambda)$), a more robust detection with regard to noise can be achieved. This is due to the deviations of $a'(\lambda)$ further deteriorating the received ratio. If only $b(\lambda)$ is modulated and detected, $a'(\lambda)$ can be set to take a fixed value and instead be used for equalization [10]. Similarly, the deviation of the received eigenvalue from the fixed, transmitted eigenvalue can be taken into account for equalization. This way, the impact of noise on the transmission can be reduced significantly, even for a single soliton receiver. This can be further expanded to correlations between eigenvalues and NFT coefficients of different channels.

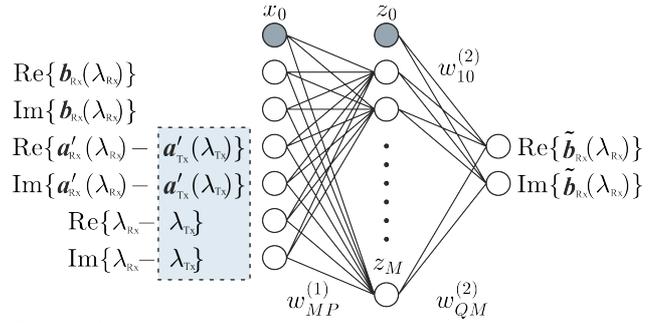

Fig. 4. Net layout with input layer (left), hidden layer and output layer (right) represented as nodes. Additionally, bias parameters $x_0$ and $z_0$ were used. The grey box highlights the known (fixed) transmission parameters.

The second considered distortion is the interference between solitons of different channels (ICI). Due to different group velocities, the solitons of different channels collide during propagation. If the desired transmission distance is close to a transmission distance at which the solitons overlap with each other in the time domain, interferences between channels occur in the linear Fourier domain and impair reception of the data. This may appear to be in contradiction with the properties of the NFT, in that eigenvalues and NFT coefficients are conserved throughout fiber propagation irrespectively of time-domain pulse collisions, and can thus be ideally retrieved. However, one should consider that an exact NFT can only be obtained if all the channels are transformed jointly. In the receiver scheme investigated here, the four channels are rather first demultiplexed and subsequently routed to one out of four coherent single-channel receivers. Moreover, the NFT analysis is done in a truncated, finite time window corresponding to the single channel pulse-to-pulse repetition time. These deviate from ideal NFT-reception and are the source of the sensitivity to pulse collisions observed in Section VI. Since the fundamental solitons are packed very closely in the linear frequency domain, they cannot be filtered out sufficiently at the receiver to fully isolate a channel. Relatively narrow optical filtering implemented at the receiver for noise reduction (Section VI) is problematic at fiber lengths coinciding with pulse collisions, due to the aforementioned pulse frequency shifts (Section II) occurring in the linear Fourier domain due to XPM, that can be up to several GHz. Since the optical filters are centered on the nominal carrier frequencies as generated at the transmitter, this leads to increased, asymmetric clipping of the pulses' Fourier content. Moreover, due to lumped amplification, a small portion of these frequency shifts remain even after collision, shifting the eigenvalues associated to the pulses [4]. These remanent frequency shifts would not occur in an ideal fiber link described by the integrable nonlinear Schrödinger equation, in which pulse power is assumed to be constant [2]. However, in the case of lumped amplification, as evaluated here, pulse power cyclically varies in between each amplifier, so that cross-phase modulation (XPM) may have asymmetric magnitudes towards the beginning and towards the end of the collision, leading to a net shift in frequency.

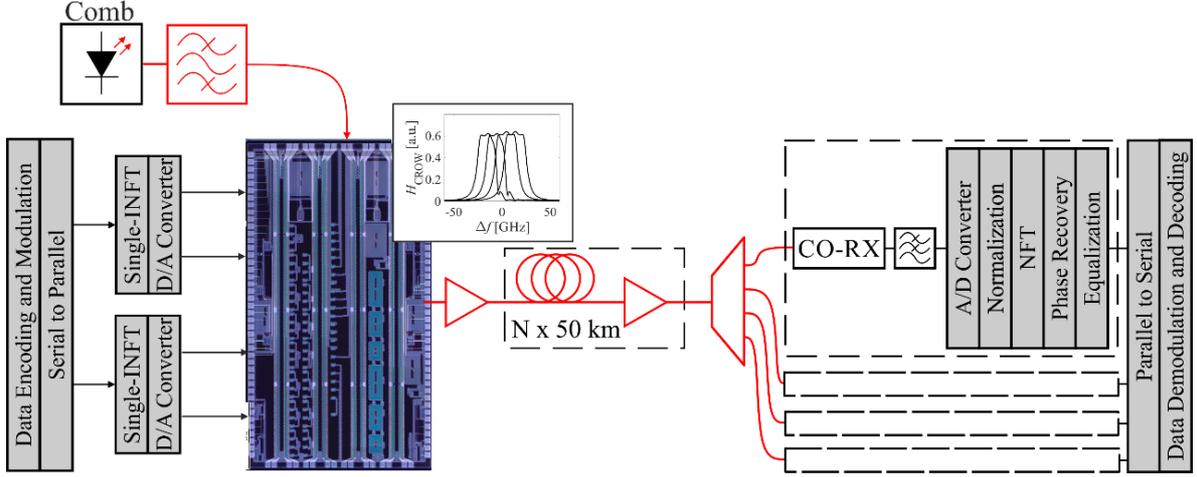

Fig. 5. Simulation setup including PIC micrograph. The insert shows the assumed PIC transfer functions.

Finally, the third distortion is the deterioration of eigenvalues and NFT coefficients due to very dense multiplexing at the transmitter, as already described in Section III. Since these deteriorations are determined by the neighboring channels, it is expected that equalization in a multi-receiver setup can tackle this.

To correct the aforementioned distortions, two approaches are investigated, a minimum mean square error (MMSE) equalizer and a neural network.

### A. MMSE approach

A simple linear minimum mean square estimate of the amplitude and phase errors $\Delta A_b$, $\Delta \phi_b$ of $b_{Rx}(\lambda)$ determined by using $\mathbf{n} = [\Delta a'_R\ \Delta a'_I\ \Delta \lambda_R\ \Delta \lambda_I]$ can improve the detection performance [10]. Here, $\Delta a' = a'_{Tx}(\lambda) - a'_{Rx}(\lambda)$ is the deviation of the fixed NFT coefficient $a'(\lambda)$ and $\Delta \lambda = \lambda_{Tx} - \lambda_{Rx}$ is the deviation of the (fixed) transmitted eigenvalue, split into real and imaginary parts since the MMSE is real valued, as indicated by subscripts $R$ and $I$. We calculate vectors $\mathbf{c}, \mathbf{d}$ that minimize the amplitude and phase deviations according to

$$\text{argmin}_\mathbf{c} \mathbf{E}[(\Delta A_b - \mathbf{c}^T \mathbf{n})^2] \\ \text{argmin}_\mathbf{d} \mathbf{E}[(\Delta \phi_b - \mathbf{d}^T \mathbf{n})^2] \quad (4)$$

where $\mathbf{E}$ denotes the expectation value [10]. $\mathbf{c}, \mathbf{d}$ are given by

$$\mathbf{c}^T = \mathbf{E}[\Delta A_b \mathbf{n}^T] \cdot \text{cov}(\mathbf{n})^{-1} \\ \mathbf{d}^T = \mathbf{E}[\Delta \phi_b \mathbf{n}^T] \cdot \text{cov}(\mathbf{n})^{-1} \quad (5)$$

This method can be expanded to multiple eigenvalues for higher order solitons.

### B. Neural network equalizer

Another approach to equalize nonlinear effects is given by a feed-forward neural network (NN) [11]. The NN used for single channel equalization in this work consists of three layers, namely the input layer, hidden layer, and output layer. Here, the input layer is initialized with $\mathbf{n}$ and the real and imaginary parts of $b_{Rx}(\lambda)$. This is depicted in Fig. 4 [11]. As for the MMSE, the NN is set up to use real-valued numbers only. 100 nodes were used inside the hidden layer. These were set up to use a sigmoid activation function on the sum of the weighted inputs, with an additional bias $x_0$ provided by the input layer. The output nodes used linear activation functions on the weighted hidden layer outputs together with a bias $z_0$. In the performed simulations, the weights were trained using the Levenberg-Marquardt algorithm [12] using training solitons. Again, the input and output layers can be expanded to multiple channels. In doing so, a neural network is capable of equalizing all channels at once taking channel interdependences into account.

## V. PIC DESIGN AND LINK BUDGET

To enable dense multiplexing and tight control of the solitons' parameters, an integrated transmitter has been developed relying on a silicon photonics PIC, based on previous experience with WDM systems [13]. By first multiplexing even and odd channels onto different bus waveguides and then combining these with a wavelength insensitive 3-dB coupler, this PIC enables in particular distortion-less superposition of fundamental soliton pulses even in case of overlapping linear spectra for nearest neighbor channels. In addition, it allows modulation of four semi-independent optical channels with two electrical data streams and two electro-optic IQ modulators, by applying each electrical channel to two optical carriers and optically delaying one of them to decorrelate the data. A micrograph of the system chip is shown in Fig. 5.

To improve the phase stability between multiple carriers, a comb source (Pilot Photonics Lyra-OCS-1000) is used. The comb source delivers a power of -6 dBm per carrier with a linewidth of approximately 80 kHz. To select four of its lines for this transmission setup, a 45 GHz passband optical filter with 2 dB insertion loss (IL) is used. To reach an appropriate injection power of 10 dBm into the PIC, as limited by grating coupler damage thresholds and silicon waveguide nonlinearities (in particular two-photon absorption), an erbium doped fiber amplifier (EDFA) with a noise figure of 5 dB is used. Carriers are coupled into the chip using a grating coupler (GC) with an IL of 3 dB. Inside the PIC, four 2nd-order coupled (ring-)resonator optical waveguide (CROW) optical add-drop multiplexers (OADMs) with an IL of 1.6 dB and a -3 dB bandwidth (BW) of 7.5 GHz route the carriers to one of two IQ-Mach-Zehnder modulators (IQ-MZMs), that each modulate two channels respectively injected and extracted from the IQ-MZMs via complementary input and output ports [14]. The IQ-MZMs have been designed with slotted transmission lines and inter-phase-shifter cross-talk suppression [15],[16]. They introduce an insertion loss and modulation penalty of 13.5 dB when operated with 1 $V_{pp}$ signals and have an electro-optic bandwidth of 14 GHz. With this configuration, channels 1 and 2, respectively, carry the same information as channels 3 and 4, but are routed out to complementary output ports of the IQ-MZMs. To enable generation of all possible signal combinations, channels 1 and 2 are then optically delayed (IL = 3 dB), thus emulating independent channels.

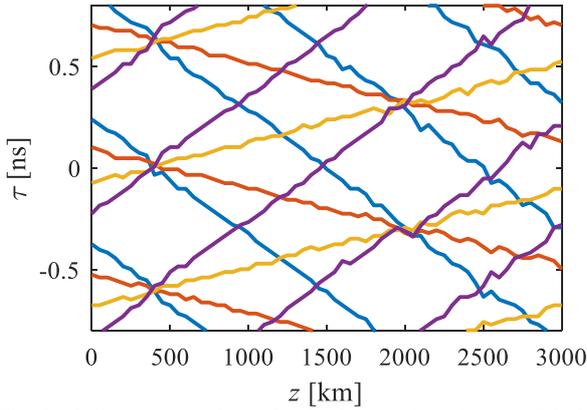

Fig. 6. Soliton peak positions for the 4 channels (color coded) inside a retarded-time window for simulation 2 at a 1550 nm center wavelength.

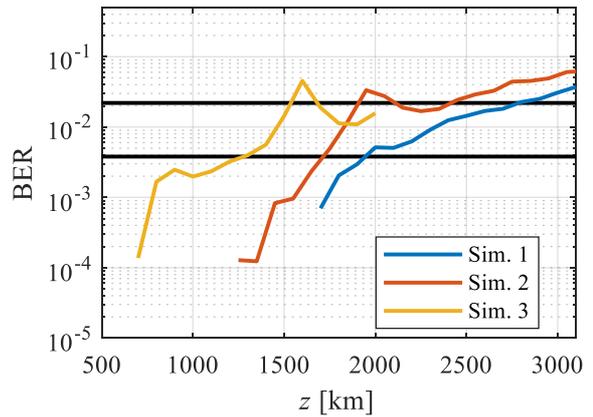

Fig. 7. Pre-FEC BERs for the three simulations without equalizers. Black lines indicate SD-FEC and HD-FEC limits at 2.2e-2 and 3.8e-3.

After modulation, channels are multiplexed onto one out of two bus waveguides using 4th-order CROW OADMs implementing Chebyshev type I filters (IL=2 dB, -3 dB BW=17.5 GHz), which can be tuned to achieve equal peak power for all channels. By multiplexing even channels onto a first bus waveguide and odd channels onto a second bus waveguide, issues related to spectral overlap of adjacent channels are avoided. Finally, a multimode interferometer coupler (MMI) combines the two bus waveguides into one, which is then routed off chip using a second GC. Due to reciprocity, the MMI results in 3 dB insertion losses, however it also results in minimal spectral distortion. Due to the aforementioned insertion losses and monitor tap losses, the resulting peak power of the solitons is -20.6 dBm off chip. Therefore, a 2nd EDFA is needed to reamplify the signal, so that launch powers can fulfill the soliton condition.

Since four fundamental solitons with different center frequencies are launched into each transmission window with equidistant timing, on-chip optical delay lines are required to shift solitons to their respective time-position in adjacent transmission windows. The delay lines are reprogrammable and enable a delay of either 300 ps or 500 ps. This allows reconfiguring the PIC for different pulse-to-pulse spacings ($\Delta T$) that can for example be chosen to be either 150 ps or 250 ps, respectively, corresponding to 4-pulse transmission windows of 600 ps or 1000 ps. Further values of $\Delta T$ are also possible with this PIC, such as the 100 ps investigated in the following. However, these also create some downtime between 4-pulse transmission windows (e.g. 100 ps for $\Delta T = 100$ ps, resulting in an overall 500 ps 4-pulse-symbol duration). Details of the PIC reconfigurability can be found in [14].

VI. TRANSMISSION SIMULATIONS

In order to investigate the impact of eigenvalue degradations due to linear de-/multiplexing of solitons and their respective eigenvalues, simulations of different transmission scenarios have been conducted.

A. Simulation setup

The simulation setup, as depicted in Fig. 5, is based on the work previously reported in [5]. To simulate different PIC settings, three simulations with different time delays between channels were set up according to Table I. This way, the symbol rates could also be changed. A total of 320,000 QPSK mapped bits are modulating $b(\lambda)$ for each single soliton stream with $\lambda = 0.5j$, with $j$ the imaginary unit. The FWHM of the solitons is set to $T_{FWHM} = 67$ ps, which leads to an electrical 99% power bandwidth of 7 GHz. After DAC conversion (6-bit vertical resolution, 20 GHz bandwidth and 92 GS/s), these solitons are modulated onto four comb lines with a carrier spacing ($\Delta f$) of 10 GHz. The normalized frequency distance (related to the real part of the eigenvalues) can then be calculated according to [4]

$$\Delta k = \pi \cdot \Delta f \cdot T_{FWHM} / 1.763, \qquad (6)$$

which results in $\Delta k = 1.2$. The remaining time-domain parameters such as $\Delta T$ and the aggregate symbol rate of a 4-soliton train ($f_{sym}$) are varied according to Table I, wherein it should be noted that the configuration of simulation 3 comprises a 100 ps down-time in between 4-soliton windows, to reflect the limited reconfigurability of the PIC, as already indicated in Section V. The bit rate is 8 times larger than $f_{sym}$ and reaches 16 Gb/s for Sim. 3, which corresponds to a spectral efficiency of 0.4 b/Hz.

TABLE I. SIMULATION PARAMETERS

|  | $\Delta T$ [$T_{FWHM}$] | $\Delta T$ [ps] | $f_{sym}$ [GBd] | $P_{launch}$ [dBm] |
|---|---|---|---|---|
| Sim. 1 | 3.74 | 250 | 1 | 2.66 |
| Sim. 2 | 2.24 | 150 | 1.666 | 4.88 |
| Sim. 3 | 1.5 | 100 | 2 | 5.67 |

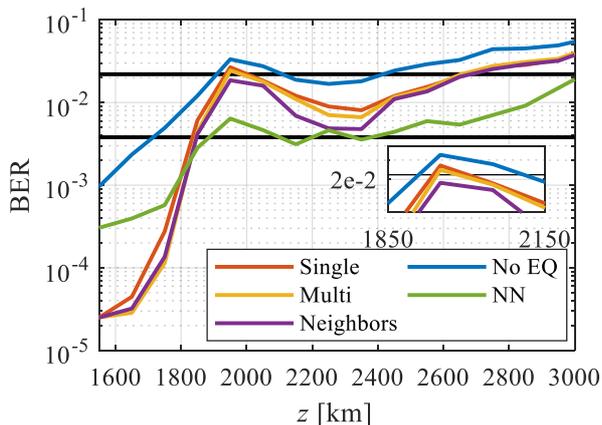
Fig. 8. Pre-FEC BERs employing different equalizers for simulation 2. Inset: BERs at the second full collision around the SD-FEC threshold.

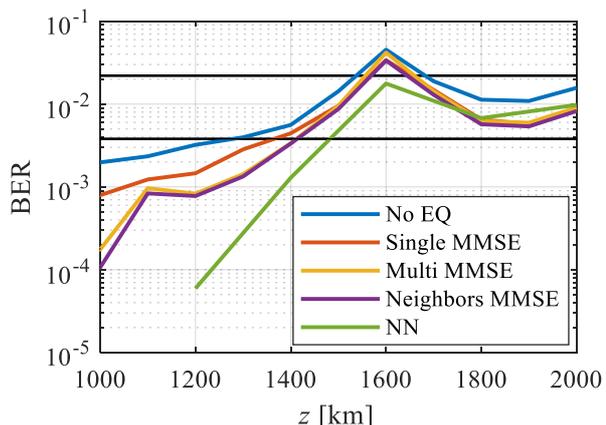
Fig. 9. Pre-FEC BERs employing different equalizers for simulation 3.

To simulate the multiplexing of the PIC, the optical solitons were attenuated to -20.6 dBm and filtered by Chebyshev filters with an optical -3 dB BW of 17.5 GHz matched to the PIC characteristics. Afterwards, an EDFA (NF = 5 dB) boosted the solitons to the launch power $P_{\text{launch}}$ indicated in Table I to reach the soliton condition for 50 km non-zero dispersion shifted fiber (NZDSF) spans ($D$ = 4.5 ps/(nm · km), $\alpha$ = 0.2 dB/km, $\gamma$ = 1.6 $W^{-1}km^{-1}$). Each span consisted of 50 km of NZDSF and one EDFA (NF = 5 dBm).

On the Rx side, four receivers were set up in parallel with local oscillator frequencies set to the transmitted carrier frequencies, without correcting for frequency shifts occurring due to soliton collisions and lumped amplification [5]-[7]. Prior to electro-optic detection, optical signals were filtered with a narrow Gaussian shaped filter with a 7 GHz -3 dB BW (3.5 GHz equivalent single-sided electrical BW) also centered on the nominal carrier frequency, to filter out as much noise as possible without deteriorating the solitons. This was followed by A/D conversion using 80 GS/s.

The DSP algorithm consisted in normalizing the solitons into NFT units, calculating the NFT using the Forward-Backward Method [17] and applying different equalization techniques as explained in Section IV.

For the MMSE equalizers, three variants were examined. The simplest setup is an independent equalization of each soliton channel according to Eqs. (4) and (5). This assumes the four receivers operating without knowledge of the neighboring channels. Moreover, preceding and following symbols are not taken into account. One step further is the equalization of each soliton also using the information of the other three receivers. Here, the solitons that were transmitted in the same time window on the other channels are taken into account. This may allow equalization of imperfections occurring due to dense multiplexing, as explained in Subsection III.B. The final setup consists of equalization using information of all four channels and with soliton information from the immediately preceding and following symbols, as defined by their launch time.

Due to different group-velocities in the channels, solitons can collide during transmission. This phenomenon is depicted in Fig. 6, in which the relative position of the solitons is shown in the time domain for simulation 2. Here, we can see that a first complete collision (collision of all 4 channels) occurs at around 450 km. A second complete collision occurs at ~2000 km. In between, single collisions between channel pairs occur.

If the receiver lies at or close to a position at which a collision occurs, the received $b(\lambda)$ and $\lambda$ can be heavily distorted, because $\Delta T$ at the receiver is close to 0, as already discussed in Section IV.

### B. Simulation results

The simulations were first conducted without any equalizer to show the penalties due to soliton collisions and eigenvalue deviations due to multiplexing. The simulated BERs without equalization are plotted in Fig. 7 for all three simulations. Here, we can clearly see the BER peak due to the full collision of the solitons of simulation 2 at 1950 km. Similar peaks can be seen for simulation 3 at 900 km and 1600 km. Furthermore, the penalty associated to post-superposition deviation of the eigenvalues, as caused by dense multiplexing, can be observed by comparing the three simulations, as error rates steadily increase with denser pulse packing. As a consequence, simulation 3 is much more limited in the achievable transmission range. If one considers the seemingly small eigenvalue deviations shown in Figs. 2 and 3 resulting from a $\Delta T$ of 1.5 $T_{\text{FWHM}}$, as assumed in simulation 3, the sensitivity of the transmission schemes to small deviations becomes apparent.

In the next step, equalizers were used to improve the results for simulations 2 and 3. Fig. 8 shows the resulting BERs after using the aforementioned equalizer approaches for simulation 2. In order to ensure enough training data, all equalizers (MMSE and NN) were trained using 10,000 training solitons. Here, all MMSE equalizers using the correlations between eigenvalue deviations and their respective NFT coefficients are highly effective. The multi-channel MMSE ("Multi" in Fig. 8), which also accounts for cross-correlations between solitons of the same transmitted 4-soliton group, is able to further improve the results of the MMSE which equalizes only one channel at a time ("Single"). This particularly applies to transmission distances up to 1800 km, i.e., before the second complete collision that also involves solitons from preceding and following launch windows not taken into account by the "Multi" MMSE. However, since in simulation 2 the solitons are multiplexed in a regime where the eigenvalues are not strongly affected by the linear superposition ($\Delta T = 2.24\, T_{\text{FWHM}}$), this effect is relatively minor and less pronounced than in simulation 3 (discussed below).

After the first collision with solitons of neighboring symbol groups occurs at ~2000 km (see Fig. 6), MMSE approaches are less effective in compensating the deviations induced by this collision. To improve on the performance of the "Multi" MMSE, the solitons of the neighboring symbol groups, that have been launched right before or after the analyzed group, can also be taken into account ("Neighbors" in Fig. 8), since pulses from these launch windows are also involved in the collision and play a role in the range 1850 – 2450 km. However, resulting further improvements are rather small. The MMSE equalizers manage to keep the BER below the SD-FEC limit of 2.2e-2 up to 2700 km.

The neural network equalizer (NN) utilized in this scenario is a single-channel equalizer only (similar to the "Single" MMSE). As seen before in [11], this equalizer performs worse compared to the MMSE equalizers for low transmission distances up to 1800 km. Afterwards, the NN approach is better suited to handle the disturbances due to collisions. Furthermore, it appears to be better capable of using the correlations between eigenvalues and NFT coefficients for long transmission distances in the low SNR regime, as it continues to significantly outperform the MMSE equalizers. The NN equalizer in this scenario is able to keep the BER close to the HD-FEC limit of 3.8e-3 after the collision, up to a total distance of 2400 km, and below the already mentioned SD-FEC limit until 3000 km. Thus, an increase in reach of up to 600 km compared to a transmission without equalizer can be achieved, if SD-FEC is assumed.

Figure 9 shows the results of the employed equalizers for simulation 3. Here, a larger "Multi" MMSE performance increase can be observed compared to a "Single" MMSE. This might be due to a partial equalization of the eigenvalue deviations arising from linear multiplexing described in Section III, that are more pronounced here due to closer spectral packing. Again, the "Neighbors'" MMSE performs best in the group of MMSEs during the collision.

The neural network equalizer is able to keep the transmission error free up to 1200 km and maintains a BER below the SD-FEC limit during collision.

## VII. Conclusion

We have presented a way to linearly multiplex solitons employing an integrated silicon photonics DWDM soliton transmitter and the nonlinear Fourier-transform. The transmitter is currently being used in experimental system tests and can modulate four first-order solitons and multiplex them using different frequency and time spacings. Transmission simulations have been conducted in accordance with the chip's characteristics. Possibilities to equalize the received signals after long-haul transmission have also been investigated and benchmarked.

Using the silicon photonics transmitter parameters in simulations, we have shown that four solitons can be merged together to be transmitted in a 500 ps window up to 2000 km using SD-FEC. If a window of 600 ps is assumed, the reach can be expanded up to 2400 km. The reach can be further improved by employing equalizers. These equalizers are able to compensate for soliton collisions, (de)multiplexing penalties and noise interactions. Using a single-channel neural network equalizer, the reach of a 600 ps window can be extended up to 3000 km.

Acknowledgment

This work was supported by the Deutsche Forschungsgemeinschaft (DFG) in the framework of the priority program "Electronic-Photonic Integrated Systems for Ultrafast Signal Processing" by grants PA 1705/1-1 and WI 4137/9-1. We also gratefully acknowledge the support of NVIDIA Corporation for the donation of the Titan Xp GPU used for this research.